%% file: reduced.tex
\documentclass[fleqn, rnote]{aa}

\usepackage{amsmath}
\usepackage{natbib}
\usepackage{psfrag, color, epsfig, rotating, pstricks, pst-node}
\usepackage{dsfont}
\usepackage{txfonts}
\usepackage{mathtools}

\bibpunct{(}{)}{;}{a}{}{,}

\newlength{\hsizethird}
\newcommand{\dd}{{\mathrm d}}

\newcommand{\ii}{{\rm i}}
\newcommand{\ee}{{\rm e}}

\newcommand{\Omegam}{\Omega_{\rm m}}

\begin{document}


\DeclareGraphicsExtensions{.eps, .ps}

\title{Fitting formulae of the reduced-shear power spectrum for weak
  lensing}
\titlerunning{Fitting formulae of the reduced-shear power spectrum}

\author{Martin Kilbinger\inst{1,2,3}}

\institute{
  Excellence Cluster Universe, Technische Universit\"at M\"unchen,
  Boltzmannstr. 2, 85748 Garching, Germany
  \and
  Universit\"ats-Sternwarte M\"unchen, Scheinerstr. 1, 81679 M\"unchen, Germany
  \and
  Shanghai Key Lab for Astrophysics, Shanghai Normal University,
  Shanghai 200234, P.~R.~China
}

\offprints{Martin Kilbinger, \email{martin.kilbinger@universe-cluster.de}}

\date{Received / Accepted}

\abstract%
%
{%
   Weak gravitational lensing is a powerful probe of large-scale
   structure and cosmology. Most commonly, second-order correlations of observed galaxy
   ellipticities are expressed as a projection of the matter power
   spectrum, corresponding to the lowest-order approximation between
   the projected and the three-dimensional power spectrum.
}%
{%
  The dominant lensing-only contribution beyond the zero-order
  approximation is the reduced shear, which takes into account not
  only lensing-induced distortions but also the isotropic
  magnification of galaxy images. This involves an integral over the
  matter bispectrum. We provide a fast and general way to calculate
  this correction term.
}%
{%
  Using a model for the matter bispectrum, we fit elementary functions
  to the reduced-shear contribution and its derivatives with
  respect to cosmological parameters. The dependence on cosmology is
  encompassed in a Taylor-expansion around a fiducial model.
}%
{%
  Within a region in parameter space comprising the WMAP7 68\%
  error ellipsoid, the total reduced-shear power spectrum (shear
  plus fitted reduced-shear correction) is accurate to 1\% (2\%) for $\ell<10^4$
  ($\ell<2 \times 10^5$). This corresponds to a factor of four
  reduction in the
  bias compared to the case where no correction is used. This
  precision is necessary to match the accuracy of current non-linear
  power spectrum predictions from numerical simulations.
}%
{%
}%

\keywords{cosmology --
gravitational lensing -- large-scale structure of the Universe}

\maketitle

\section{Introduction}
\label{sec:intro}

Cosmic shear surveys exploit the distortions of distant galaxy images
induced by the matter structures in the Universe on large scales. Weak
cosmological lensing has become an important tool to measure
cosmological parameters. The current precision in the measured shapes
of high-redshift galaxies is at the few percent level providing
important constraints on the matter density $\Omegam$ and the
power-spectrum normalisation $\sigma_8$. Future surveys aim to achieve
sub-percent level accuracy in measuring dark-energy properties or
deviations from general relativity.

To reach this precision for cosmological parameters, efforts not only
on the observational but also on the theoretical side have to be
made. To predict the shear power spectrum from theoretical models,
several approximations are usually made to facilitate
calculations. When comparing and fitting models to observations, these
simplifications can significantly bias the resulting cosmological
parameters. One of these approximations is to neglect the correction
due to the so-called reduced shear.

The observed ellipticity of galaxies is an estimator of the
reduced shear
\begin{equation}
  g = \frac{\gamma}{1 - \kappa},
  \label{g_red_1}
\end{equation}
where $\gamma$ is the shear and $\kappa$ is the convergence. Here and
in the following, both shear and reduced shear are represented as
complex quantities, $\gamma = \gamma_1 + \ii \gamma_2$, $g = g_1 + \ii
g_2$. Since in weak lensing the convergence is small, $|\kappa| \ll
1$, one usually simplifies the above relation to linear order, $g
\approx \gamma$.
 
The effect of reduced shear was examined using $N$-body
simulations by \citet{W05} and compared with theoretical predictions
in \cite{2006PhRvD..73b3009D}. Corrections to the lensing skewness
were discussed in \cite{1998MNRAS.296..873S} and
\cite{2005PhRvD..72h3001D}.
An extensive study of
higher-order corrections to the weak-lensing power spectrum, including
reduced shear, was presented by \cite{2009arXiv0910.3786K}. Compared
with corrections due to light-path integration (relaxation of the Born
approximation) and lens-lens coupling
\citep[e.g.,][]{1993LIACo..31..579S,1997A&A...322....1B}, reduced-shear
corrections represent the dominant contribution to the weak-lensing
power spectrum. The lowest-order correction term, which is of order
${\cal O}(\phi^3)$ in the gravitational potential $\phi$, reaches 10\%
of the total lensing spectrum. This is the term that we consider
in this work. The next-order correction at ${\cal O}(\phi^4)$ is one
magnitude below the ${\cal O}(\phi^3)$-term \citep{2009arXiv0910.3786K}.
Ignoring reduced shear will cause a bias in cosmological parameters,
which is comparable to the statistical error from future surveys like
DES\footnote{http://www.darkenergysurvey.org}, LSST\footnote{http://www.lsst.org},
or Euclid\footnote{http://www.ias.u-psud.fr/imEuclid}
\citep{2009ApJ...696..775S}. For the weak-lensing results of
the COSMOS survey\footnote{http://cosmos.astro.caltech.edu}, an estimate from
numerical simulations yielded a 1\% underestimation of $\sigma_8$
\citep{SHJKS09}.

Another correction to weak lensing from a magnitude-limited galaxy
sample is the so-called \emph{magnification
  bias}. \cite{2001MNRAS.326..326H} found its contribution to the
lensing power spectrum to be small. The corresponding third-order
correction terms were also calculated and compared to numerical
simulations by \cite{2003AA...403..817M}. However, if not corrected for,
future surveys will yield significantly biased cosmological parameter
constraints \citep{2009ApJ...702..593S}.  The lowest-order correction
term has the same functional form as the reduced-shear one, and
therefore both can be dealt with simultaneously. Higher-order terms
contribute by smaller than one percent to the lensing power
spectrum \citep{2009arXiv0910.3786K}.

There are additional important sources of contamination to the weak
lensing observables, some of which may be larger than the
reduced-shear corrections. These involve galaxy clustering, the
intrinsic alignment of galaxies with each other or with the
surrounding dark matter haloes, and the modelling of the non-linear dark-matter
power spectrum and baryonic processes. The modelling of these effects
requires detailed understanding of baryonic physics, galaxy formation,
the interplay between galaxies and dark matter, and massive numerical
simulations in the case of the non-linear power spectrum.
In contrast, taking into account the reduced shear is straightforward
and can be done with sufficient accuracy to render it a sub-dominant
contamination.

In the next section, we review the expressions for the lowest-order
reduced-shear term, and introduce a fitting scheme to efficiently
model this term. In Sect.~\ref{sec:res}, we present our results,
demonstrating the accuracy of the fits. We conclude this paper
in Sect.~\ref{sec:disc}.

\section{Reduced shear}
\label{sec:g}

\subsection{Lowest-order expansion}

For small values of the convergence, $|\kappa| < 1$, the reduced
  shear (\ref{g_red_1}) is expanded as
\begin{equation}
  g = \frac{\gamma}{1-\kappa} = \gamma [ 1 + \kappa
    + \kappa^2
    + \ldots ] .
  \label{g_red}
\end{equation}
To calculate the power spectrum of the reduced shear $g$, we apply
a Fourier transform to Eq.~(\ref{g_red}) to second order
\begin{equation}
  \hat g(\vec \ell) \approx \hat \gamma(\vec \ell) + (\hat \gamma * \hat
  \kappa)(\vec \ell) .
\end{equation}
 In our notation, the convolution of $\hat
\gamma$ and $\hat \kappa$ is
\begin{equation}
  (\hat \gamma * \hat \kappa)(\vec \ell) = \int
  \frac{\dd^2\ell^\prime}{(2\pi)^2} \hat \gamma(\vec \ell^\prime) \hat
  \kappa(\vec \ell - \vec \ell^\prime).
  \label{faltung}
\end{equation}
We assume that the reduced shear is measured in redshift bins. With
$P_g^{mn}$, we denote the reduced-shear (cross-)power spectrum for two
redshift bins with indices $m$ and $n$. In general, the shear field
can be decomposed into an E-mode (gradient) and a B-mode
(curl). However, as is the case for cosmic shear at lowest order, the
reduced-shear correction of order ${\cal O}(\phi^3)$ does not have a
B-mode component \citep{2002A&A...389..729S}, and in the following we
focus only on the E-mode.  In the absence of a B-mode, the
reduced-shear power spectrum is defined to be
\citep{2002A&A...389..729S}
\begin{align}
  \left\langle \hat g^m(\vec \ell) {\hat g^n}{^*}(\vec \ell^\prime) \right\rangle
  = \; & (2\pi)^2
  \delta_{\rm D}(\vec \ell - \vec \ell^\prime) P^{mn}_{g}(\ell),
  \label{def-Pg}
\end{align}
where $g^m$ denotes the reduced shear measured from galaxies in the
$m^{\rm th}$ redshift bin, the asterisk ($^*$) denotes complex
conjugation, and $\delta_{\rm D}$ is the Dirac delta-function.

The reduced-shear power spectrum is the sum of the convergence power
spectrum (which equals the shear power spectrum) and a correction
given by
\begin{align}
  P^{mn}_{g}(\ell) & \approx P^{mn}_{\kappa}(\ell) +
  {P^{(1)}}^{mn}_{g}(\ell);
  \label{Pgtot} \\
  (2\pi)^2 \delta_{\rm D}(\vec \ell -
  \vec \ell^\prime) {P^{(1)}}^{mn}_{g}(\ell) & = 
 \left\langle \hat \gamma^m(\vec \ell)
      {\left( \hat \gamma^n * \hat
          \kappa^n \right)}^*(\vec \ell^\prime) \right\rangle +
    (m \leftrightarrow n).
 \label{Pg1E}
\end{align}
Using the relation between shear and convergence, $\hat
\gamma(\vec \ell) = \exp(2\ii\beta) \hat \kappa(\vec \ell)$, where
$\beta$ is the polar angle of the wave vector $\vec \ell$, and that
the convergence is a real field, $\hat \kappa^*(\vec \ell) = \hat
\kappa(-\vec \ell)$, the third-order correlator in Eq.~(\ref{Pg1E})
can be written in terms of the convergence bispectrum $B_\kappa$,
which is defined by the equation
\begin{align}
  \left\langle \hat \kappa^m(\vec \ell_1) \hat \kappa^n(\vec \ell_2) \hat
      \kappa^p(\vec \ell_3) \right\rangle = 
    (2\pi)^2 \delta_{\rm D}(\vec \ell_1 + \vec \ell_2 + \vec \ell_3)
    B^{mnp}_\kappa(\vec \ell_1, \vec \ell_2).
  \label{def-Bk}
\end{align}
With this, we derive the correction term
\begin{align}
{P^{(1)}}^{mn}_{g}(\ell) = \int
  \frac{\dd^2\ell^{\prime}}{(2\pi)^2}
 \ee^{-2\ii\beta^{\prime}}
 \left[ B^{mnn}_\kappa(\vec \ell_0, -\vec \ell^{\prime}) +
   B^{nmm}_\kappa(\vec \ell_0, -\vec \ell^{\prime}) \right].
 \label{gsgk}
\end{align}
Without loss of generality, we have set $\beta = 0$ and introduced $\vec
\ell_0 = (\ell, 0)$. In the integrand, $\beta^\prime$ is the polar
angle of $\vec \ell^\prime$.

The convergence bispectrum can be expressed in terms of the three-dimensional matter
bispectrum $B_\delta$ using Limber's equation 
\citep{1953ApJ...117..134L,1992ApJ...388..272K} providing
\begin{align}
  B_\kappa^{mnp}(\vec \ell_1, \vec \ell_2) = & \int_{a_{\rm lim}}^1
  \dd a \, F^{nmp}(a) \,
 B_\delta\left(
    \frac{\vec \ell_1}{f_K[\chi(a)]},
    \frac{\vec \ell_2}{f_K[\chi(a)]}
      ; \chi(a)
  \right); \notag \\
  F^{mnp}(a) = & \frac{\dd \chi}{\dd a}
  \frac{G^m(a)G^n(a)G^p(a)}{f_K[\chi(a)]} .
  \label{Bkappa}
\end{align}
The lower limit of the integral over the scale factor $a$ corresponds
to the limiting redshift, $a_{\rm lim} = 1/(1+z_{\rm lim})$, $\chi$
is the comoving distance, and $f_K$ is the comoving angular diameter distance.
The lensing efficiency $G^m$ for the $m^{\rm th}$ redshift bin is
\begin{equation}
  G^m(a) = \frac 3 2 \left(\frac{H_0}c\right)^2 \frac{\Omegam}{a}
  \int_{a_{\rm lim}}^a \dd a^\prime p_m(a^\prime) \frac{f_K[\chi(a^\prime) - \chi(a)]}{f_K[\chi(a^\prime)]},
\end{equation}
which includes an integral over the probability distribution
$p_m$ of the number density of source galaxies in the $m^{\rm th}$
bin.

\subsection{Expansion around a fiducial model}

In the expressions needed to calculate the first-order reduced-shear power spectrum
in Eqs.~(\ref{gsgk}, \ref{Bkappa}), we can interchange the $\ell$- and the
$a$-integration, and replace the latter by a finite sum. Then
\begin{align}
  {P^{(1)}}^{mn}_{g}(\ell) & \approx \Delta a \sum_{k=1}^{N_a}
  \bar F^{mn}(a_k) Q(\ell, a_k);
  \notag \\
  \bar F^{mn}(a) & = \frac 1 2 \left[ F^{mnn}(a_k)  + F^{nmm}(a_k) \right];
  \notag \\
  Q(\ell, a) & = 2 \int \frac{\dd^2 \ell^\prime}{(2\pi)^2} \cos (2 \beta^\prime)
  B_\delta\left(\frac{\vec\ell_0}{f_K[\chi(a)]},
    \frac{-\vec\ell^\prime}{f_K[\chi(a)]}\right).
  \label{sum_a}
\end{align}
Evaluating the $\ell^\prime$-integral is not a difficult but very
time-consuming numerical problem. It is significantly slower than the
calculation of the matter power spectrum necessary for the convergence
power spectrum. This makes reduced-shear corrections the limiting
factor in the parameter analysis of a weak-lensing survey, for example
using Monte Carlo sampling. In the following, we present fitting
formulae that provide an accurate approximation for this term within
a region in parameter space around a fiducial cosmological model. The
$\ell$-independent term $F^{mnn}(a)$ is easily computed and can be
tabulated. Since it depends on the source redshift distribution, it
has to be determined for each survey.

We perform a Taylor-expansion to first order ${P^{(1)}}^{mn}_{g}$ as a function of a cosmological
parameter vector $\vec p = (p_1, p_2, \ldots p_M)$ around a fiducial cosmological model with
parameter $\vec p_0$
\begin{align}
  {P^{(1)}}^{mn}_{g} (\ell, \vec p) \approx {P^{(1)}}^{mn}_{g} (\ell, \vec p_0) + (\vec p
  - \vec p_0) \left. \vec \nabla_{\vec p} {P^{(1)}}^{mn}_{g} (\ell, \vec
    p)\right|_{\vec p = \vec p_0}. 
  \label{Pg_Taylor}
\end{align}
Inserting Eq.~\eqref{sum_a}, the derivatives with respect to cosmological
parameters are given by
\begin{align}
  \frac{\partial {P^{(1)}}^{mn}_{g}(\ell, \vec p)}{\partial
      p_\alpha} =
    \Delta a
    \sum_{k=1}^{N_a} \Bigg\{ &
    \bar F^{mn}(a_k, \vec p) \frac{\partial Q(\ell, a_k, \vec
      p)}{\partial p_\alpha}
  \notag \\
  & +
    \frac{\partial \bar F^{mn}(a_k, \vec p)}{\partial p_\alpha} Q(\ell, a_k,
    \vec p) \Bigg\}.
  \label{dP_dp}
\end{align}
In Sect.~\ref{sec:res}, we consider the cosmological
parameters $\vec p = (\Omega_{\rm m}, \Omega_{\rm de}, w, \Omega_{\rm
  b}, h, \sigma_8, n_{\rm s})$ (see also Table \ref{tab:limits}).

\subsection{Magnification and size bias}

A typical galaxy sample used in weak cosmological lensing is selected
by both magnitude and galaxy size. Magnification due to lensing
changes both quantities \citep[e.g.][]{BS01}, and therefore introduces
a correlation between number density and convergence. If the number
density of galaxies with fluxes higher than some flux $S$, and sizes
larger than some size $R$ can be written as power laws,
\begin{equation}
  p(>S, >R) \propto S^{-\alpha} R^{-\beta},
\end{equation}
the observed number density differs from the unlensed one $p_0$
to first order, becoming
\begin{equation}
  p = p_0 \left( 1 + q \kappa \right);
  \qquad q = 2 \left( \alpha + \beta - 1 \right),
\end{equation}
according to \citet{2009ApJ...702..593S,2009PhRvL.103e1301S}.
This simple linear model assumes that the galaxy selection function
depends on flux and size in a separable way. We refer to
\citet{2003ApJ...583...58W} for a generalisation that takes into account
correlations between the two quantities.

Magnification and size bias induces a lowest-order correction to the
lensing power spectrum which has the same form as for reduced
shear. Therefore, we can add the corresponding correction term
${P^{(1)}}^{mn}_{\mu}$ to Eq.~(\ref{Pgtot}) with
\begin{align}
  {P^{(1)}}^{mn}_\mu(\ell) & \approx \Delta a \sum_{k=1}^{N_a}
  \bar F^{mn}(a_k) Q_\mu(\ell, a_k);
  \notag \\
  Q_\mu(\ell, a) & = q Q(\ell, a),
\end{align}
where the function $Q$ is given in Eq.~(\ref{sum_a}).

\subsection{Fitting formulae}

For simplicity, we define $Q^{(0)} = Q$ and $Q^{(\alpha)} = \partial
Q/\partial p_\alpha$ for $\alpha = 1 \ldots M$.
These correspond to the $\ell$-dependent terms in
Eqs.~(\ref{sum_a}) and (\ref{dP_dp}), which behave as near-power laws for
both small and large $\ell$. With the abbreviation $y=\ln \ell$, we
perform two linear fits of $\ln\left| Q^{(\alpha)}(y)\right|$ for $y<y_{\rm low} = 3$ and
$y>y_{\rm up}=11.5$, respectively.
We smoothly piece together these two asymptotic functions with a cubic
spline $s^{\alpha}(y)$ such that the composite function $\ln h^{(\alpha)}(y), \alpha
= 0 \ldots M,$ and its first derivative are continuous
\begin{align}
  \ln h^{(\alpha)}(y, a)  = &
  \left\{
    \renewcommand{\arraystretch}{1.3}
      \begin{array}{ll}
        g_{\rm low}^{(\alpha)}(y, a) \quad & \mbox{for} \quad
        y<y_{\rm low} \\
        s^{(\alpha)}(y, a) & \mbox{for} \quad y_{\rm
          low}\le y \le y_{\rm up} \;\; ; \\
        g_{\rm up}^{(\alpha)}(y, a)  \quad & \mbox{for} \quad
        y > y_{\rm up} \\
      \end{array}
    \renewcommand{\arraystretch}{1}
    \right.
    \label{h-composite}
\end{align}
for the linear functions
\begin{align}
  g_{\rm low}^{(\alpha)}(y, a)  & = b^{(\alpha)}_0(a) \cdot y +
  b^{(\alpha)}_1(a); \notag \\
  g_{\rm up}^{(\alpha)}(y, a)  & = b^{(\alpha)}_2(a) \cdot y+
  b^{(\alpha)}_3(a),
\end{align}
and the cubic spline
\begin{align}
  s^{(\alpha)}(y, a) = & \; \sum_{\nu=0}^{3} r^{(\alpha)}_\nu(a)
  \cdot \left(\frac{y -
      y_{\rm low}}{\Delta}\right)^\nu ; \notag \\
  \Delta = & \; y_{\rm up} - y_{\rm low}; \notag \\
  r^{(\alpha)}_0(a)  = & \; g^{(\alpha)}_{\rm low}(y_{\rm low}, a); \notag \\
  r^{(\alpha)}_1(a)  = & \; \Delta \cdot b^{(\alpha)}_0(a); \notag \\
  r^{(\alpha)}_2(a)  = & \; 3 g^{(\alpha)}_{\rm up}(y_{\rm up}, a) - \Delta \cdot b^{(\alpha)}_2(a) - 3
                        r^{(\alpha)}_0(a) - 2 r^{(\alpha)}_1(a) ; \notag \\
  r^{(\alpha)}_3(a)  = & \; \Delta \cdot b^{(\alpha)}_2(a) - 2
  g^{(\alpha)}_{\rm up}(y_{\rm up}, a)
  + 2 r^{(\alpha)}_0(a) + r^{(\alpha)}_1(a) .
\end{align}
The ratios $ Q^{(\alpha)} / h^{(\alpha)}$ are then fitted
by polynomials of order $N$ in $y = \ln \ell$, at the fiducial
cosmology $\vec p = \vec p_0$, and for each scale factor $a = a_k, k=1 \ldots N_a$,
\begin{align}
  Q^{(\alpha)}(y,a)
 & \approx h^{(\alpha)}(y, a) \cdot \sum_{i=0}^{N} c^{(\alpha)}_i(a) \, y^i;
  \;\;\;\; \alpha = 0 \ldots M.
  \label{fit-QdQ}
\end{align}
We note that we cannot fit the logarithm of the functions $Q^{(\alpha)}$,
since the derivatives of $Q$ with respect to some cosmological
parameters change sign.

The fit coefficients $b^{(\alpha)}_i, c^{(\alpha)}_i$ are expected to
smoothly vary with the scale factor $a$. We therefore perform another series
of fits by polynomials of order $N_b$ and $N_c$, respectively
\begin{align}
  b^{(\alpha)}_i(a) \approx \sum_{j=0}^{N_b} B^{(\alpha)}_{ij} \, a^j; 
  \qquad
  c^{(\alpha)}_i(a) \approx \sum_{j=0}^{N_c} C^{(\alpha)}_{ij} \, a^j.
 \label{fit-coeff}
\end{align}
The two matrices $B^{(\alpha)}$ and $C^{(\alpha)}$ for $\alpha = 0
\ldots M$ defined in the last equation completely determine our
approximation for the $\ell$-dependent terms of the reduced-shear
power spectrum correction.

\section{Results}
\label{sec:res}

We use a fiducial flat $\Lambda$CDM cosmological model with WMAP7-like
parameters, $\Omegam = 0.27, \Omega_{\rm b } = 0.045, h = 0.71,
\sigma_8 = 0.8$ and $n_{\rm s} = 0.96$
\citep{2010arXiv1001.4538K}. The dark-matter bispectrum is calculated
according to \cite{2001MNRAS.325.1312S}. Their fitting formula is
accurate to only 30\% - 50\% on small scales; this is however
sufficient for our purpose. We note also that the bispectrum is not
calibrated for any dark-energy model other than $\Lambda$CDM. For the
matter power spectrum, we use the `halofit' fitting formula of
\cite{2003MNRAS.341.1311S} and the transfer function `shape fit'
from \cite{1998ApJ...496..605E}. Following the \texttt{icosmo.org} code
\citep{2008arXiv0810.1285R} for models with $w \ne -1$, we
modify `halofit' to interpolate between $\Lambda$CDM and $w=-1/3$,
which behaves in a similar way to an OCDM model \citep[for
more details, see][]{SHJKS09}.

\begin{figure}[!tb]
  
  \resizebox{\hsize}{!}{
    \includegraphics{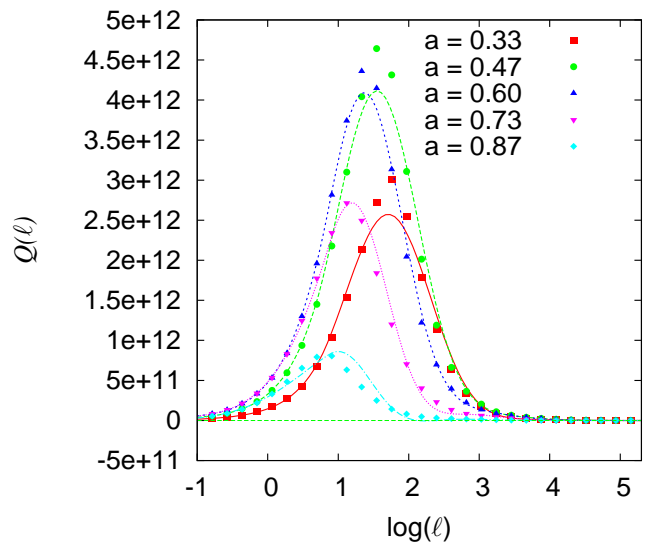}
  }
  
  \caption{The function $Q^{(0)}(\ell) = Q(\ell)$ for different scale factors
    $a$ as indicated in the panel (points), and the composite fitting
    function (Eq.~\ref{fit-QdQ}, lines).}
  \label{fig:QdQ}
\end{figure}

\begin{figure*}[!tb]
  
  \resizebox{\hsize}{!}{
    \includegraphics{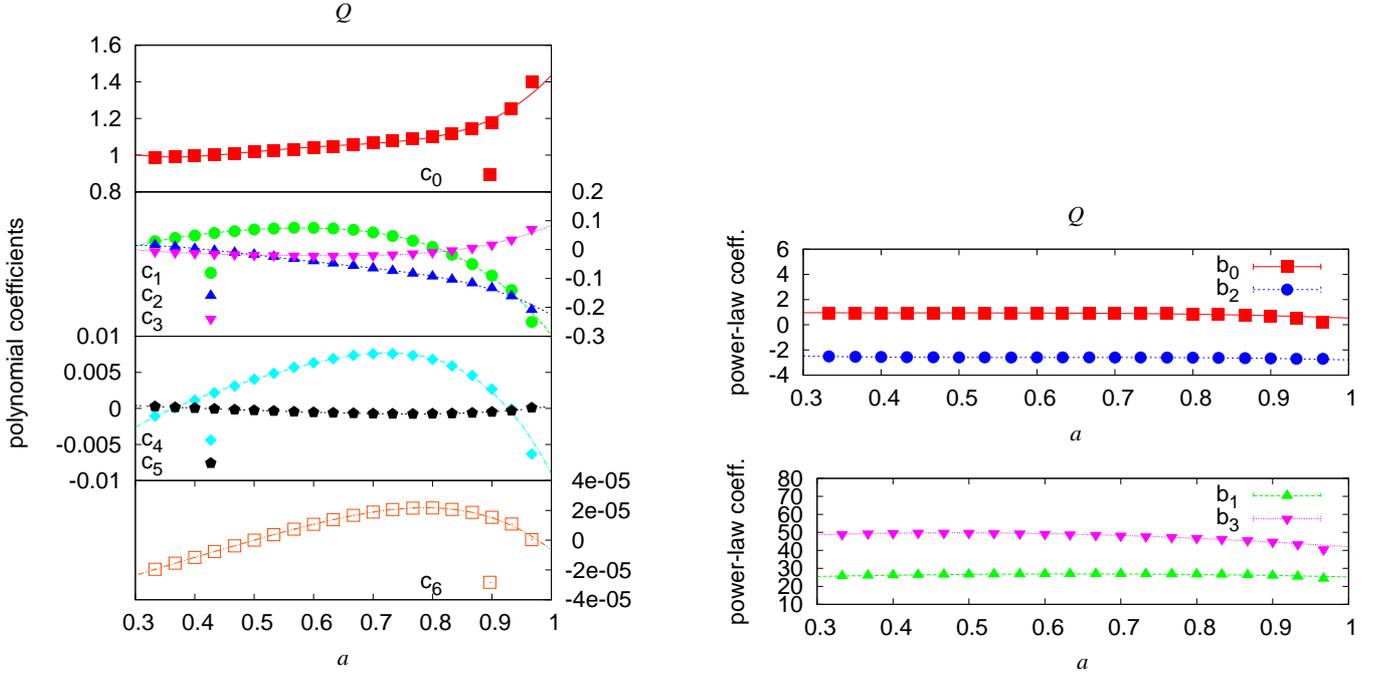}
  }
  
  \caption{Fit coefficients as a function of scale factor.
    \emph{Left:} The polynomial fit coefficients $c^{(0)}_i$ for $i=0
    \ldots N=6$ (Eq.~\ref{fit-QdQ}). The lines are polynomials in $a$
    or order $N_c = 3$.
    \emph{Right:} The power-law coefficients $b^{(0)}_i, i=0 \ldots 3$ of
    the composite function $h^{(0)}$ (Eq.~\ref{h-composite}). The
    lines are polynomials in $a$ of order $N_b = 3$.}
  \label{fig:coeffs}
\end{figure*}

The function $Q$ and the corresponding fit with the composite function
given in Eq.~(\ref{fit-QdQ}) are shown in Fig.~\ref{fig:QdQ}. The polynomial that
is part of the composite function has order $N=6$.  The fitting
coefficients $b^{(0)}_i$ and $c_i^{(0)}$ are plotted in
Fig.~\ref{fig:coeffs}. Although the higher-order polynomial
coefficients have relatively low amplitudes (right panel), we found that a
polynomial of order 6 is necessary to provide a good fit to the
reduced-shear power spectrum, as discussed below.  For the
polynomial fits of the coefficients as functions of $a$ (\ref{fit-coeff}), we chose
$N_b = N_c = 3$. These
cubic polynomials provide sufficient accuracy, in particular
for $a \rightarrow 1$, where the coefficients show the most
variation. This is important because the reduced-shear correction
spectrum in Eq.~(\ref{sum_a}) obtains a large contribution from large
$a$.

We perform the fits in the $\ell$-range between $0.1$ and $2 \times
10^5$. The functions $Q^{(\alpha)}$ are not perfect power laws,
therefore the fit for large $\ell$ is not excellent. We found an
improvement of our final results for the total power spectrum by
adding 0.05 to $b_2$ after performing the fits.

The fitting functions for the reduced-shear power spectrum corrections
provide accurate results over a wide range in $\ell$. We illustrate
the case of a single redshift bin with distribution
$p(z) = (z/z_0)^\alpha \cdot \exp[-(z/z_0)\,^\beta]$ and parameters
$\alpha = 2, \beta = 1.5$, and $z_0 = 0.5$. The maximum considered redshift is
$z_{\rm max} = 2$, which results in a mean redshift of 0.75.
At $\ell = 10^3$ the reduced-shear correction to the
convergence power spectrum starts to become important. On smaller
scales, the latter underestimates the total power spectrum by more
than 1\%. For $10^3 < \ell < 10^5$, we fit $P^{(1)}_{\rm g}$
(Eq.~\ref{Pg1E}) to better than 20\%. This is sufficient to provide an
approximation of the total power spectrum at the percent-level.

\begin{figure*}[!tb]
  
  \resizebox{\hsize}{!}{
    \includegraphics{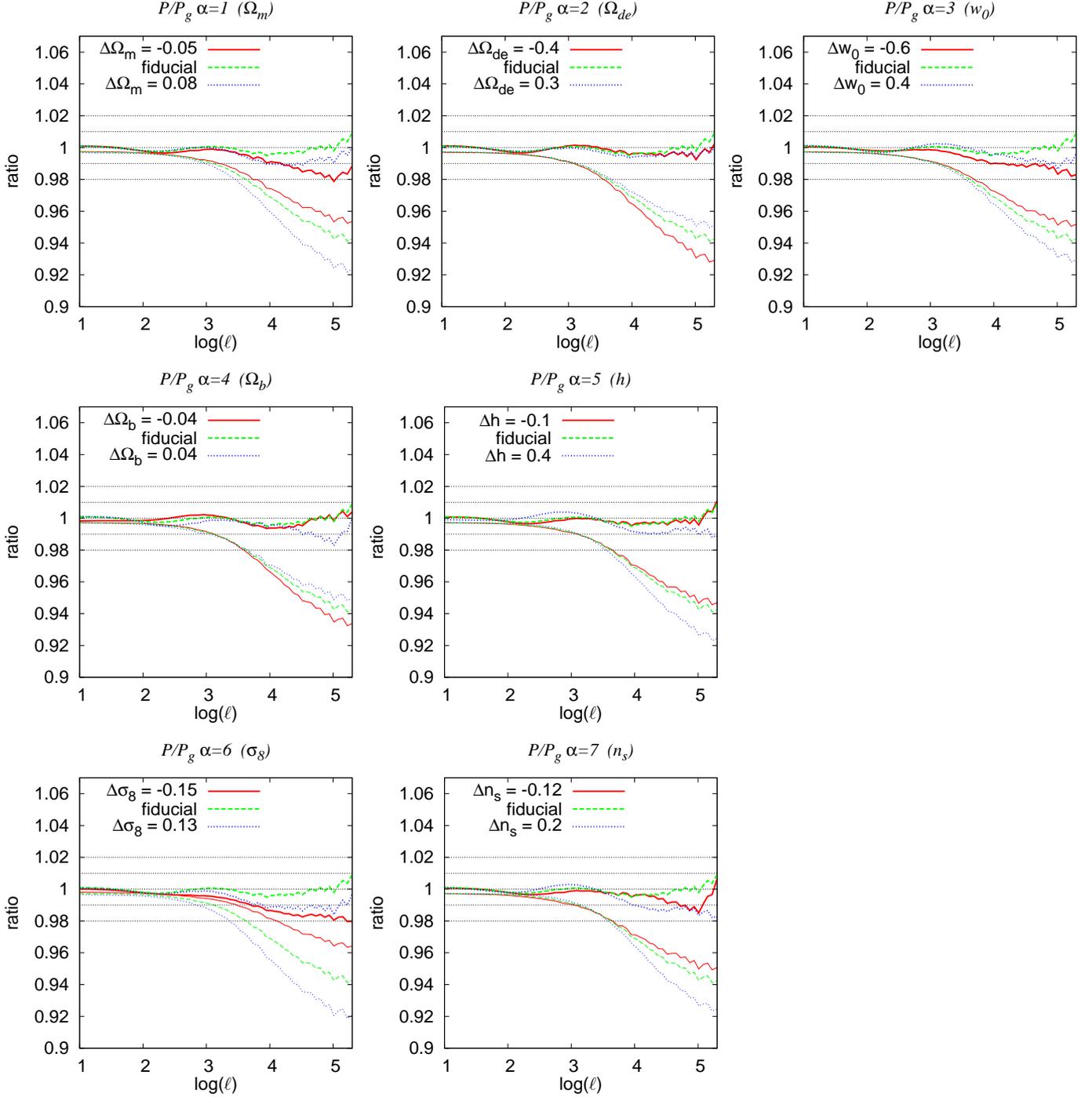}
  }
  
  \caption{The fitted reduced-shear power spectrum (thick lines) and
    the uncorrected convergence power spectrum (thin lines). Both
    quantities are plotted as ratios with respect to the reduced-shear
    power spectrum (\ref{Pgtot}), where the first-order correction
    $P^{(1)}_{\rm g}$ has been obtained by numerical integration of
    Eq.~(\ref{sum_a}). In each panel, we show the power spectra at the
    fiducial model (dashed lines) and the lower (solid) and upper
    (dashed) boundaries within which a 2\% accuracy is reached for
    $\ell < 2 \times 10^5$, see Table \ref{tab:limits}.}
  \label{fig:Pg_Pkx}
\end{figure*}

In Fig.~\ref{fig:Pg_Pkx}, we the plot the ratio of the reduced-shear power
spectrum derived using our fitting functions to that obtained by numerical
integration of Eq.~(\ref{sum_a}). This is compared to the case of no
correction for reduced-shear, corresponding to just the convergence
power spectrum. In this latter power spectrum, a downward bias is evident, since
the power is underestimated. This bias increases from 1\% at
$\ell = 1000$ to 6\% at $\ell = 10^5$ for the fiducial
cosmology. In models with more structure, this bias is larger,
e.g., 8\% for $\sigma_8 = 0.93$. In contrast, our fitted correction is
accurate to better than 1\%  for $\ell = 2 \times 10^5$ at the fiducial
point.

We test different redshift distributions by changing to $z_0 = 0.3$ and $z_0 =
0.7$, corresponding to mean redshifts of 0.45 and 1, respectively, and
also a redshift bin of width 0.1 around z=0.75. In all cases, the
fitting formula remains accurate to within 1\% at the fidicual model.

In Table \ref{tab:limits}, we denote the parameter ranges for which our
fitting formulae are accurate to better than $1\%$ for $\ell<10^4$,
and better than $2\%$ for $\ell<2 \times 10^5$.

\begin{table}

 \caption{Parameter limits for where the accuracy of the fitted
    reduced-shear power spectrum is better than $1\%$ for $\ell<10^4$,
    and better than $2\%$ for $\ell<2 \times 10^5$, see also
    Fig.~\ref{fig:Pg_Pkx}. Note that for each given range, the other
    parameters were kept fixed.}
  \label{tab:limits}

  \input limits_a

\end{table}

\section{Conclusions}
\label{sec:disc}

The lowest-order correction term for the reduced-shear power spectrum
is the dominant contribution from lensing-only effects
\citep{2009arXiv0910.3786K}. This term is proportional to the third
power of the gravitational potential $\phi$, and includes an integral
over the lensing bispectrum. In this paper, we have presented fitting
formulae for this integral and its derivative with respect to
cosmological parameters. This has permitted a more efficient
calculation of the reduced-shear correction power spectrum within a
parameter range comprising the 68\% confidence region of
WMAP7. The fitting scheme reduces the computational effort from tens
of minutes for the full integration to a fraction of a second.

For currently available weak-lensing data, the uncertainty in the
non-linear power spectrum on small scales is still larger than the
bias due to the reduced-shear. For example, the widely-used fitting
prescription by \citet{2003MNRAS.341.1311S} strongly underestimates
the power on small scales.
More recent numerical simulations however provide fitting formulae
that improve the accuracy by a factor 5 to 10 \citep{CoyoteI,
  CoyoteII, CoyoteIII}. Moreover, their emulation scheme can be
applied to a wide range of cosmological parameters. This range is
comprised in the hypercube for which our fits are
valid. In combination, these results provide us with predictions for the
weak lensing power spectrum that are accurate to the few percent level.

\section*{Acknowledgments}

We thank Peter Schneider for helpful comments on the manuscript, and
the anonymous referee whose suggestions helped to improve the
paper. This research was supported by the DFG cluster of excellence “Origin
and Structure of the Universe” and
the Chinese National Science
Foundation Nos. 10878003 \& 10778725, 973 Program No. 2007CB 815402,
Shanghai Science Foundations and Leading Academic Discipline Project
of Shanghai Normal University (DZL805).

\begin{appendix}

\section{Fitting matrices}

The matrices $B^{(\alpha)}$ and $C^{(\alpha)}$ (\ref{fit-coeff}), $a = 0
\ldots M$ contain the coefficients of the reduced-power
spectrum fits given in Eq.~(\ref{fit-QdQ}). Here, we provide the numerical values from
our fits. The index $\alpha=0$ corresponds to the function $Q$
(\ref{sum_a}) in the fiducial cosmology, and $\alpha=1 \ldots 7$ to its
derivatives with respect to cosmological parameters (see Table
\ref{tab:limits}). The matrices are also available in electronic form
with an example code\footnote{http://www2.iap.fr/users/kilbinge/reduced}.

{\small
\input B_in
\input C_in
}

\end{appendix}

\bibliographystyle{aa} \bibliography{astro}

\end{document}

%% file: limits_a.tex
\begin{center}
\begin{tabular}{|l|l|l|l|}\hline
\rule[-3mm]{0em}{8mm}$\alpha$	 &Parameter	 &lower	 &upper	\\\hline\hline
1	 &$\Omega_\textrm{m}$	 &0.22	 &0.35	\\\hline
2	 &$\Omega_\textrm{de}$	 &0.33	 &1.03	\\\hline
3	 &$w$	 &-1.6	 &-0.6	\\\hline
4	 &$\Omega_\textrm{b}$	 &0.005	 &0.085	\\\hline
5	 &$h$	 &0.61	 &1.11	\\\hline
6	 &$\sigma_8$	 &0.65	 &0.93	\\\hline
7	 &$n_\textrm{s}$	 &0.86	 &1.16	\\\hline
\end{tabular}
\end{center}

%% file: B_in.tex
\begin{align}
B^{(0)} & = \left(
\begin{array}{llll}
1.2157	 &-1.7061	 &3.613	 &-2.588	\\
22.054	 &14.525	 &-10.134	 &-1.2063	\\
-1.7221	 &-4.2894	 &6.9843	 &-3.7669	\\
41.719	 &34.891	 &-40.957	 &6.3019	\\
\end{array}
\right)
;\notag
\end{align}
\begin{align}
B^{(1)} & = \left(
\begin{array}{llll}
1.2963	 &-2.2435	 &4.7574	 &-3.4183	\\
24.802	 &15.438	 &-12.793	 &0.30538	\\
-0.42199	 &-14.769	 &28.968	 &-16.938	\\
62.035	 &-22.702	 &-10.113	 &15.12	\\
\end{array}
\right)
;\notag
\end{align}
\begin{align}
B^{(2)} & = \left(
\begin{array}{llll}
1.2314	 &-1.826	 &3.9322	 &-2.9291	\\
27.286	 &-8.0841	 &21.999	 &-18.426	\\
-3.2808	 &4.3912	 &-6.464	 &2.8446	\\
62.447	 &-73.471	 &122.62	 &-71.766	\\
\end{array}
\right)
;\notag
\end{align}
\begin{align}
B^{(3)} & = \left(
\begin{array}{llll}
1.1772	 &-1.523	 &3.3478	 &-2.6269	\\
17.16	 &39.066	 &-56.363	 &24.476	\\
1.7985	 &-17.673	 &25.431	 &-12.429	\\
-1.2271	 &192.08	 &-259.8	 &113.95	\\
\end{array}
\right)
;\notag
\end{align}
\begin{align}
B^{(4)} & = \left(
\begin{array}{llll}
1.1983	 &-1.5973	 &3.3941	 &-2.4472	\\
21.942	 &28.174	 &-32.269	 &10.746	\\
-4.4314	 &9.5715	 &-14.092	 &6.4373	\\
75.097	 &-124.48	 &200.67	 &-109.85	\\
\end{array}
\right)
;\notag
\end{align}
\begin{align}
B^{(5)} & = \left(
\begin{array}{llll}
1.3643	 &-2.6937	 &5.6748	 &-4.0482	\\
22.681	 &17.078	 &-13.713	 &0.12838	\\
-1.6597	 &-3.5025	 &5.6229	 &-3.2007	\\
44.397	 &11.923	 &-3.5699	 &-10.617	\\
\end{array}
\right)
;\notag
\end{align}
\begin{align}
B^{(6)} & = \left(
\begin{array}{llll}
1.2149	 &-1.7012	 &3.6033	 &-2.5827	\\
24.779	 &8.4417	 &-0.27702	 &-6.3675	\\
-2.5782	 &-0.084018	 &0.53439	 &-0.56922	\\
53.073	 &-11.638	 &28.888	 &-27.294	\\
\end{array}
\right)
;\notag
\end{align}
\begin{align}
B^{(7)} & = \left(
\begin{array}{llll}
1.3469	 &-3.3556	 &6.6862	 &-4.5164	\\
25.294	 &6.7075	 &4.1411	 &-10.121	\\
-1.0794	 &-6.1604	 &9.2899	 &-4.7514	\\
37.736	 &49.648	 &-61.197	 &16.44	\\
\end{array}
\right)
.\notag
\end{align}

%% file: C_in.tex
\begin{align}
C^{(0)} & = \left(
\begin{array}{llll}
0.56428	 &2.3001	 &-3.9649	 &2.427	\\
0.12548	 &-0.94677	 &2.59	 &-2.04	\\
0.1557	 &-0.65321	 &0.89795	 &-0.58724	\\
-0.063989	 &0.43324	 &-1.0211	 &0.7215	\\
0.0087569	 &-0.093102	 &0.24571	 &-0.1693	\\
-0.00055709	 &0.008454	 &-0.022712	 &0.015138	\\
1.4809\cdot10^{-5}	 &-0.00027429	 &0.00072501	 &-0.00046762	\\
\end{array}
\right)
;\notag
\end{align}
\begin{align}
C^{(1)} & = \left(
\begin{array}{llll}
-0.43501	 &-3.1459	 &5.6111	 &-3.4686	\\
-0.087793	 &0.80516	 &-2.761	 &2.4422	\\
-0.10227	 &0.33267	 &-0.18646	 &0.1432	\\
0.073105	 &-0.5126	 &1.2891	 &-0.95999	\\
-0.023766	 &0.19417	 &-0.48231	 &0.33371	\\
0.0030621	 &-0.024925	 &0.059059	 &-0.039071	\\
-0.0001238	 &0.00098714	 &-0.0022652	 &0.0014607	\\
\end{array}
\right)
;\notag
\end{align}
\begin{align}
C^{(2)} & = \left(
\begin{array}{llll}
-0.49063	 &-2.7133	 &4.5927	 &-2.7656	\\
0.13912	 &-0.76984	 &0.7525	 &0.11512	\\
-0.22771	 &1.0278	 &-1.3875	 &0.79182	\\
-0.00013136	 &-0.029954	 &0.26883	 &-0.30505	\\
0.01479	 &-0.046285	 &-0.0031371	 &0.04091	\\
-0.0020694	 &0.0066385	 &-0.0028282	 &-0.0018872	\\
8.0142\cdot10^{-5}	 &-0.00026039	 &0.00016485	 &1.1648\cdot10^{-5}	\\
\end{array}
\right)
;\notag
\end{align}
\begin{align}
C^{(3)} & = \left(
\begin{array}{llll}
0.25229	 &4.0227	 &-6.6135	 &3.7371	\\
-0.52388	 &3.3599	 &-5.5903	 &2.5675	\\
0.35958	 &-1.7899	 &2.4962	 &-1.27	\\
0.076671	 &-0.50746	 &0.7332	 &-0.2437	\\
-0.045615	 &0.2431	 &-0.33587	 &0.13382	\\
0.0052982	 &-0.026584	 &0.03616	 &-0.014855	\\
-0.00018401	 &0.0008918	 &-0.0012	 &0.00049882	\\
\end{array}
\right)
;\notag
\end{align}
\begin{align}
C^{(4)} & = \left(
\begin{array}{llll}
0.12913	 &4.7128	 &-8.0056	 &4.5759	\\
0.13963	 &-0.11394	 &0.61465	 &-0.97735	\\
0.26098	 &-1.3324	 &1.9954	 &-1.1216	\\
-0.11282	 &0.48918	 &-0.99176	 &0.70867	\\
0.016542	 &-0.075864	 &0.19759	 &-0.1533	\\
-0.0010708	 &0.0055907	 &-0.017359	 &0.0139	\\
2.6106\cdot10^{-5}	 &-0.00015916	 &0.00055278	 &-0.00044773	\\
\end{array}
\right)
;\notag
\end{align}
\begin{align}
C^{(5)} & = \left(
\begin{array}{llll}
-0.31944	 &-3.8608	 &6.7541	 &-4.0259	\\
-0.010529	 &-0.63367	 &1.0269	 &-0.067878	\\
-0.24657	 &1.3721	 &-2.1149	 &1.1977	\\
0.051632	 &-0.14509	 &0.34221	 &-0.34002	\\
0.0010064	 &-0.030306	 &0.0014399	 &0.042272	\\
-0.00070106	 &0.0052506	 &-0.0026698	 &-0.0029094	\\
3.3959\cdot10^{-5}	 &-0.00020532	 &0.00010945	 &8.9968\cdot10^{-5}	\\
\end{array}
\right)
;\notag
\end{align}
\begin{align}
C^{(6)} & = \left(
\begin{array}{llll}
0.58241	 &2.2178	 &-3.8312	 &2.3591	\\
0.15574	 &-1.1301	 &3.0374	 &-2.3593	\\
0.1333	 &-0.53689	 &0.68283	 &-0.46414	\\
-0.073256	 &0.48689	 &-1.1436	 &0.80567	\\
0.012694	 &-0.11623	 &0.29589	 &-0.20194	\\
-0.00098447	 &0.011061	 &-0.028339	 &0.018735	\\
2.9427\cdot10^{-5}	 &-0.00036705	 &0.00092519	 &-0.0005942	\\
\end{array}
\right)
;\notag
\end{align}
\begin{align}
C^{(7)} & = \left(
\begin{array}{llll}
-0.45031	 &-3.1685	 &5.4862	 &-3.2435	\\
0.088613	 &-0.95773	 &1.3824	 &-0.2107	\\
-0.20975	 &1.2013	 &-1.8198	 &1.0069	\\
0.022754	 &-0.030319	 &0.17275	 &-0.24886	\\
0.0067345	 &-0.052022	 &0.030858	 &0.028623	\\
-0.0011801	 &0.0070072	 &-0.004861	 &-0.0020472	\\
4.8691\cdot10^{-5}	 &-0.00025807	 &0.0001707	 &6.9607\cdot10^{-5}	\\
\end{array}
\right)
.\notag
\end{align}